\documentclass[aps,prl,twocolumn]{revtex4}
\usepackage{graphicx,epsf}

\newcommand{\lbco}{La$_{2-x}$Ba$_x$CuO$_4$}

\begin{document}

\title {
Magnetic excitations in a bond-centered stripe phase: \\
Spin waves far from the semi-classical limit
}

\author{Matthias Vojta and Tobias Ulbricht}
\affiliation{\mbox{Institut f\"ur Theorie der Kondensierten Materie,
Universit\"at Karlsruhe, 76128 Karlsruhe, Germany}}
\date{June 7, 2004}

\begin{abstract}
Using a spin-only model, we compute spin excitation spectra
in a bond-centered stripe state with long-range magnetic order.
We employ a bond operator formalism, which naturally captures
both dimerization and broken spin symmetry in a unified framework.
At low energies, the spin excitations resemble spin waves,
but at higher energies they are very similar to spin-1 excitations
of isolated spin ladders.
Our theory does well describe neutron scattering data
[J. M. Tranquada {\em et al.}, Nature {\bf 429}, 534 (2004)]
on \lbco, pointing towards bond order in this material.
\end{abstract}
\pacs{74.72.-h,75.10.Jm}

\maketitle


For a number of cuprate superconductors it has been established,
most notably by neutron scattering techniques \cite{lsco,mook,waki},
that states with incommensurate spin and charge correlations,
commonly referred to as stripes,
appear over a significant range of the phase diagram.
While in some materials these correlations remain dynamic \cite{ssrmp,stevek},
in others they become static and apparently coexist with
superconductivity at lowest temperatures.
The role of these stripes for superconductivity has been discussed
extensively \cite{ek,pnas,jan,doug,bondsite},
but is at present not fully understood.

Recent experiments \cite{jt04} have mapped out the spin excitations
in an ordered stripe phase of \lbco\ at a hole doping of $x=\frac 1 8$.
In this paper we will present a consistent theoretical description of these
neutron scattering data for all energies.

For cuprates with dopings near $x=\frac 1 8$ the spatial period of the charge order
is found to be four lattice spacings. Furthermore,
the period of the spin modulation equals twice the charge modulation
period, i.e., the ordering wavevectors obey ${\bf K}_c = 2 {\bf K}_s$.
The microscopic structure of stripes has not yet been fully elucidated.
Regarding the symmetry of the charge modulation w.r.t. reflection on a
Cu-Cu bond axis one distinguishes bond-centered and site-centered
stripes \cite{bondsite}.
Theoretical proposals based on doping a paramagnetic Mott insulator
predict the coexistence of superconductivity with bond order \cite{ssrmp,vs},
compatible with bond-centered stripes.
On the other hand, theories starting from the ordered antiferromagnet
and favoring site-centered stripes have also been put forward.
For states with spin and charge stripe order it has been suggested
that the magnetism in the hole-poor regions resembles that of the undoped
antiferromagnet, with local $(\pi,\pi)$ ordering;
then the stripes act as antiphase domain walls in the magnetic
background.

The neutron scattering spectra on \lbco\ \cite{jt04}
indicate a linearly dispersing mode near the ordering wavevector at low energies,
not inconsistent with a spin-wave picture.
However, at higher energies conventional spin waves
(as have been employed in Refs.~\onlinecite{scheidl,erica} to stripe-ordered phases)
do not appear to describe the data,
which instead are closer to what one expects for gapped two-leg ladders.

\begin{figure}
\epsfxsize=2.9in
\centerline{\epsffile{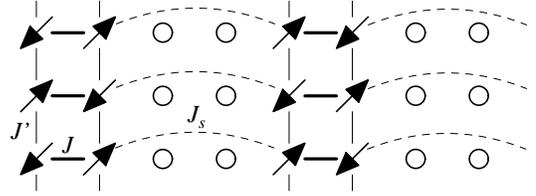}}
\caption{
Sketch of the bond-centered stripe state considered here.
We use a two-dimensional (2d) spin-only model, where the effect of the hole-rich stripes
is to mediate a weak coupling, $J_s<0$, between hole-poor regions,
accross the hole-rich ones. The hole-poor regions are spin ladders,
characterized by couplings $J,J'$.
The unit cell consists of 8 Cu sites, i.e., 4 spins in the effective
Heisenberg model.
\vspace*{-10pt}
}
\label{fig:geom}
\end{figure}

The purpose of this paper is to point out that there exists a simple
unified description of the spin dynamics at all energies.
The crucial point is the coexistence of dimerization and {\em weak} magnetic
order in bond-centered stripes.
The spin ordering in the hole-poor region is far from the semi-classical
limit (this is the limit described by conventional spin waves),
but instead is better understood as being close to a quantum phase transition
where magnetic order disappears.
We capture this physics using a spin-only model for the hole-poor regions,
assuming that the hole-rich stripes provide a weak coupling between
them.
For simplicity, we focus on stripe states with charge-order period of four
sites, where both the hole-rich and hole-poor regions have width two.
Thus we consider a Heisenberg model with a geometry shown in
Fig.~\ref{fig:geom}, with couplings $J$, $J'$, $J_s$ for
rungs, legs, and across the stripes, respectively.
$J$ and $J'$ are antiferromagnetic, whereas $J_s$ is much weaker
and ferromagnetic, mimicking the antiphase property of the stripe.
This model has a quantum paramagnetic singlet phase and
a magnetically ordered phase with ordering wavevector ${\bf K}_s = (3\pi/4,\pi)$
in units of the Cu square lattice.
The transition will occur at a critical $J_s^{cr}$ which depends on $J,J'$.
For $|J_s|$ of order $J,J'$ one expects a well-ordered state with negligible
dimerization where excitations can be described by conventional spin waves,
however, here we are interested in $J_s$ closer to the critical value,
where quantum effects are strong and spin-wave theory needs to be modified.


{\it Bond operator formalism.}
A natural language for spin dimerization is the bond operator representation
for spins \cite{SaBha}.
For two spins $1/2$ spins $\vec{S}_{1}$, $\vec{S}_{2}$ one introduces bosonic operators
for creation of a singlet
$s^{\dagger}|0\rangle = \frac{1}{\sqrt{2}}(|\uparrow \downarrow\rangle - |\downarrow \uparrow\rangle)$
and three triplet states $t_{\alpha}^{\dagger}, \alpha = x,y,z$  above a fictitious
vacuum $|0\rangle$:
$t_{x}^{\dagger}|0\rangle =-\frac{1}{\sqrt{2}} (|\uparrow \uparrow\rangle - |\downarrow \downarrow\rangle)$,
$t_{y}^{\dagger}|0\rangle = \frac{i}{\sqrt{2}}(|\uparrow \uparrow\rangle +|\downarrow \downarrow\rangle)$,
$t_{z}^{\dagger}|0\rangle = \frac{1}{\sqrt{2}}(|\uparrow\downarrow\rangle + |\downarrow \uparrow\rangle)$.
Then the following representation is exact:
$
S_{1,2}^{\alpha} = \frac{1}{2} ( \pm s^{\dagger}  t_{\alpha} \pm
t_{\alpha}^{\dagger} s  - i \epsilon_{\alpha\beta\gamma} t_{\beta}^{\dagger}
t_{\gamma})
$.
The physical Hilbert space is defined by the constraint
$s^{\dagger}s + t_{\alpha}^{\dagger}t_{\alpha} = 1$.
For our system, Fig.~\ref{fig:geom}, we will use bond operators
for each rung of the ladders.
The labels ``1'' (``2'') will be assigned to the left (right) spin
of each rung.

The simplest singlet ground state corresponds to a product state
of the form $|\phi_0\rangle = \prod_i s_i^{\dagger}|0\rangle$, where
$i$ labels rungs (dimers).
Magnetic ordering on top of the dimerized background corresponds
to condensation of one of the triplets.
A suitable product state is given by \cite{norice,sommer}
\begin{eqnarray}
|\tilde{\phi}_0\rangle
&\sim&
\prod_i (s^{\dagger}_i+\lambda \text{e}^{i{\bf Q'R}_i}t^{\dagger}_{iz})|0\rangle
\,.
\label{gs}
\end{eqnarray}
where $\bf Q'=(\pi,\pi)$ is the ordering wavevector on the {\em dimer} lattice.
A non-zero value of $\lambda$ corresponds to a state with magnetic order:
it has a condensation amplitude $\lambda$ of $t_z$ triplets, and
interpolates between the singlet product state ($\lambda=0$) and the
classical N\'eel state ($\lambda=1$).
The ordered state has 8 Cu sites per unit cell, which translates into
two dimers per unit cell of the spin-only model, i.e., the dimers form
a bipartite lattice with ``checkerboard'' order.
The quantity $\lambda$ can be understood as a variational parameter:
minimizing the energy of $|\tilde{\phi}_0\rangle$ with the full Hamiltonian
gives
\begin{equation}
\lambda^2 = \frac{2J'-J_s - J}{2J'-J_s + J}
\label{lameq}
\end{equation}
for $2J'-J_s>J$, and $\lambda = 0$ otherwise.
This defines the location of the transition point at the mean-field level,
we comment on corrections beyond mean-field below.

To describe fluctuations around the product state, $|\tilde{\phi}_0\rangle$,
we perform a basis transformation \cite{sommer}:
\begin{eqnarray}
\tilde{s}^{\dagger}_i &=&
  (s^{\dagger}_{i}+ \lambda \text{e}^{i\mathbf{Q'}\mathbf{R_i}}
  t^{\dagger}_{iz}) / \sqrt{1+\lambda^2}\,, \nonumber \\
\tilde{t}^{\dagger}_{iz} &=& (-\lambda \text{e}^{i\mathbf{Q'}\mathbf{R_i}}
  s^{\dagger}_{i}+ t^{\dagger}_{iz} ) / \sqrt{1+\lambda^2} \,, \nonumber \\
\tilde{t}^{\dagger}_{ix} &=& t^{\dagger}_{ix} \,,~
\tilde{t}^{\dagger}_{iy} = t^{\dagger}_{iy}
\,.
\label{op-trafo}
\end{eqnarray}
The product state (\ref{gs}) 
takes the form
$
| \tilde{\phi}_0\rangle = \prod_i \tilde{s}_i^{\dagger}\left|0\right\rangle
$.

The Hamiltonian can now be re-written in this new basis.
The lengthy expression, containing bilinear, cubic, and quartic boson terms,
can be found in Refs.~\onlinecite{sommer,tlcucl}. 
[Linear boson terms, which can appear in principle, exactly vanish
for the correct choice of $\lambda$ (\ref{lameq})].
The simplest non-trivial approach is a harmonic approximation \cite{sommer,oleg,tlcucl},
similar to the Holstein-Primakoff approach to spin waves.
It is obtained by resolving the constraint as
$\tilde{s}=\tilde{s}^\dagger=(1-\tilde{t}^\dagger_\alpha\tilde{t}_\alpha)^{1/2}$,
expanding the root, and keeping only bilinear terms in the $\tilde{t}$ bosons.
The resulting Hamiltonian is found as:
\begin{eqnarray}
\tilde{{\cal H}} &=& \sum_{{\bf q}\alpha} A_{{\bf q}\alpha}\tilde{t}^\dagger_{{\bf q}\alpha}\tilde{t}_{{\bf q}\alpha}
   + \sum_{{\bf q}\alpha} \frac{B_{{\bf q}\alpha}}{2}\left(
    \tilde{t}^\dagger_{{\bf q}\alpha}\tilde{t}^\dagger_{{\bf -q}\alpha}  + {\rm h.c.}
\right).
\label{fourier-h}
\end{eqnarray}
Here, $\tilde{t}_{{\bf q}\alpha}$ are the modified basis operators which have been Fourier
transformed with respect to the dimer momentum $\bf q$.
The functions $A_{\bf q}, B_{\bf q}$ contain all information about the
geometry of the system. For our model we have
\begin{eqnarray}
A_{{\bf q}x} &=& J \frac{1}{1+\lambda^2} + K \frac{\lambda^2}{(1+\lambda^2)^2} \nonumber \\
&& + J' \frac{1-\lambda^2}{1+\lambda^2} \cos q_y - \frac{J_s}{2}\cos q_x
,\nonumber\\
B_{{\bf q}x} &=& J' \cos q_y - \frac{J_s}{2} \frac{1-\lambda^2}{1+\lambda^2} \cos q_x
,\nonumber\\
A_{{\bf q}z} &=& J \frac{1-\lambda^2}{1+\lambda^2} + K \frac{2\lambda^2}{(1+\lambda^2)^2} + B_{{\bf q}z}
,\nonumber\\
B_{{\bf q}z} &=& \big(J' \cos q_y - \frac{J_s}{2}\cos q_x\big) \frac{(1-\lambda^2)^2}{(1+\lambda^2)^2}
\label{akbk}
\end{eqnarray}
with $K = 4 J'-2J_s$ and
$A_{{\bf q}y} = A_{{\bf q}x}$, $B_{{\bf q}y} = B_{{\bf q}x}$.
A Bogoliubov transformation of (\ref{fourier-h}) leads to
$
\tilde{{\cal H}} = \sum_{{\bf q}\alpha} \omega_{{\bf q}\alpha} \tau^\dagger_{{\bf q}\alpha}\tau_{{\bf q}\alpha}
$ 
where $\omega^2_{{\bf q}\alpha} = A_{{\bf q}\alpha}^2 - B_{{\bf q}\alpha}^2$.
In the paramagnetic state, $\lambda=0$, the excitations are threefold
degenerate and correspond to gapped triplet modes \cite{SaBha,norice,sommer,oleg,tlcucl,eder}.
As has been shown in Ref.~\onlinecite{sommer}, for $\lambda=1$ one
recovers standard linear spin-wave theory, with $\tilde{t}_x,\tilde{t}_y$ being the
transverse spin-wave modes, and $\tilde{t}_z$ representing a longitudinal mode
(which is dispersionless and carries zero spectral weight at $\lambda=1$).
Note that -- although we have two dimers per magnetic unit cell -- three magnetic
modes are obtained, which is due to the sublattice symmetry of the dimer lattice.
The formalism provides a consistent description of both zero-temperature
phases including the transition (albeit with mean-field critical exponents),
and is able to describe Goldstone modes which arise on the ordered side close
to the magnetic quantum phase transition.
It can also be extended for external magnetic fields \cite{sommer,tlcucl},
and has recently been employed to describe the quantum phase transitions
of the coupled dimer system TlCuCl$_3$ (Ref.~\onlinecite{tlcucl}).

So far, we have restricted ourselves to a harmonic approximation of independent bosons.
Clearly, this approximation overestimates the tendency to magnetic order, and
underestimates the energy of the excitations (as the hard-core repulsion is relaxed).
For the paramagnetic phase, it is possbile to incorporate the repulsion, e.g.,
with the Brueckner method \cite{kotov};
however, we do not know of a consistent way to extend such an approach to the
ordered phase with a smooth connection between the two.
Thus, we continue to work with the independent boson picture, but take into account
interaction effects by using {\em renormalized} parameters.
This idea has been studied for the single spin ladder in Ref.~\onlinecite{eder},
where it was shown that a quantitatively accurate description of the single-particle
dynamics is possible.
The most important renormalization is found in the $q=0$ contribution to the excitation
energy (the momentum-independent term in $A_{\bf q}$), which is
$J_{\rm ren} =1.77 J$ instead of $J$ (for $\lambda=0$) -- this will be used below.
(Different renormalizations for the $A_{\bf q}$ and $B_{\bf q}$ cannot
be easily taken in account, as we have to require Goldstone's theorem to be fulfilled.)


\begin{figure}[!t]
\epsfxsize=3in
\centerline{\epsffile{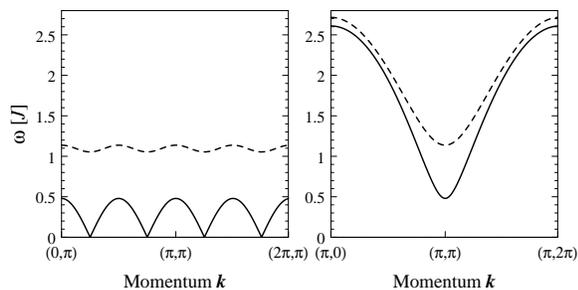}}
\caption{
Dispersions $\omega(\bf k)$ of the magnetic modes, calculated for
vertical bond-centered stripes with charge period four, see Fig.~\ref{fig:geom},
using the harmonic approximation (see text).
Solid: Doubly degenerate transverse mode.
Dashed: Longitudinal mode.
Parameter values are $J=J'=100$ meV, $J_s = -0.06 J$,
and the $J$ renormalization (see text) has been used \cite{eder,param}.
For these parameters the ``condensation'' amplitude is
$\lambda = 0.27$ and the staggered magnetization $M_s = 0.30$;
where $\lambda=1$, $M_s=0.61$ would correspond to the spin-wave result in the
semi-classical (N\'eel) limit of a square-lattice antiferromagnet.
\vspace*{-10pt}
}
\label{fig:disp}
\end{figure}

{\it Results.}
We focus on parameters $J=J'$ and small $J_s$,
and replace $J$ by $J_{\rm ren}$ in Eq. (\ref{akbk}) \cite{param,error}.
Sample dispersion relations are plotted in Fig.~\ref{fig:disp};
note that the momenta $\bf k$ in the Cu lattice and $\bf q$
in the dimer lattice are related by $k_x = q_x/4$, $k_y = q_y$.
The doubly degenerate transverse Goldstone mode and the longitudinal mode are
clearly distinguished; at higher energies, both modes resemble the dispersion
of triplets in a two-leg ladder.

Next we calculate the $T\!=\!0$ susceptibility as measured in inelastic neutron scattering.
A physical spin operator can create both single-particle and two-particle
excitations out of the ground state,
according to
$S_\alpha({\bf k}) =
[ t^\dagger_\alpha({\bf q})+t_\alpha(-{\bf q}) ] (1-e^{i k_x})/2 -
i \epsilon_{\alpha\beta\gamma} \sum_{\bf q_1} t_{\beta}^{\dagger}({\bf q}+{\bf q}_1) t_{\gamma}({\bf q}_1) (1+e^{i k_x})/2$
(where we have set $s=1$).
Thus, the magnetic response will consist of sharp peaks from elementary excitations
and a two-particle continuum.
In the following, we restrict our attention to the
one-particle contributions, as the continuum will be hard to detect
experimentally.
Averaging over the neutron spin polarizations,
we obtain $\chi''({\bf k},\omega)$ as sum of $\delta$ peaks
with weights determined by various matrix element terms.
Importantly, there appear factors of $|1-e^{i k_x}|^2$ in the
matrix elements for the one-particle processes, which enhance the
response near ${\bf k} = (\pi,\pi)$.
Also, the weight of the longitudinal mode is suppressed by a factor
of $[(1-\lambda^2)/(1+\lambda^2)]^2$.

\begin{figure}[!t]
\includegraphics[width=3.2in]{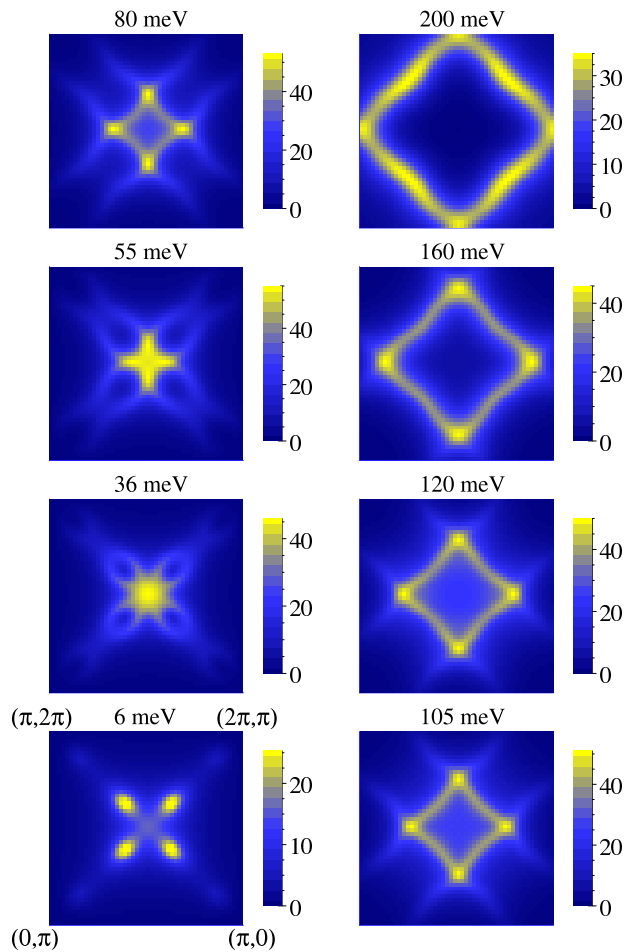}
\caption{(color)
Neutron scattering intensity, $\chi''({\bf k},\omega)$:
The panels show fixed energy cuts as function of momentum in the
{\em magnetic} Brillouin zone.
Parameter values are as in Fig.~\protect\ref{fig:disp}.
The $\delta$ peaks have been replaced by Lorentzians with width $\Gamma = 0.2 J$,
and the responses of horizontal and vertical stripes have been added.
The background arising from the two-particle continuum is not shown.
The figure can be directly compared to Fig. 2 of Ref.~\protect\onlinecite{jt04}.
\vspace*{-10pt}
}
\label{fig:res1}
\end{figure}

In order to compare with the experiment of Ref.~\onlinecite{jt04}
we add the contributions from horizontal and vertical stripes,
and plot the result as function of the external momentum
at fixed energy.
A sample result is shown in Fig.~\ref{fig:res1} where we have broadened the
$\delta$ peaks to account for the experimental resolution.
Interesting features are observed:
For small energies (e.g. 6 meV) the response consists of four peaks
representing the four cones of spin-wave modes.
With increasing energy the cones widen, however, the outer part becomes
suppressed in intensity due to matrix element effects. Thus, the
spectrum at 36 meV is dominated by the inner part of the spin-wave
cones [close to $(\pi,\pi)$].
For even higher energies (where the inter-ladder coupling becomes unimportant),
the modes gradually change their character -- towards triplet modes of a dimerized
ladder -- and the scattering intensity forms a diamond which moves outward
with increasing energy.
(The neglect of boson interactions in our approximation
somewhat overestimates the mode dispersion near the zone boundary --
this does not qualitatively influence the result for $\chi''$.
It also has to be kept in mind that the precise form of the
dispersion at high energies depends on microscopic details
like the presence of ring-exchange etc.)
Overall, one observes a good agreement with the experimental
data of Tranquada {\em et al.} \cite{jt04}.


{\it Conclusions.}
We have discussed the spin excitations in a bond-centered stripe phase,
where strong dimerization coexists with weak collinear magnetic order.
Using a bond operator approach 
we have been able to consistently describe the modes for all energies:
The transverse modes resemble semi-classical spin waves at low energies,
but cross over into triplet modes of dimerized ladders at higher energies.
The crossover energy, associated with a saddle point in the mode dispersion,
is only a fraction of $J$ for systems with weak antiferromagnetism being close
to a magnetic quantum phase transition.

Two ingredients are crucial for the ``dual'' character of the modes:
(i) strong quantum fluctuations (dimerization), and
(ii) the presence of two energy scales for the dispersion, namely
a large $J'$ and a small $J_s$.
Interestingly, a plaquette-ordered state with 2d charge modulation obeys
only (i), hence there is no crossover in the character of the
mode \cite{oleg}.
On the other hand, spin waves for stripes \cite{scheidl,erica} only
capture (ii), resulting in a high-energy spectrum different from ours.

The agreement of our description with the experiments \cite{jt04,raman}
suggests that the spin ordering in \lbco\ is far from the semi-classical limit,
and a spin-wave description may only be used at lowest energies.
Instead, strong quantum effects related to dimerization dominate the spectrum
at higher energies.
These results strongly point to the existence of bond order.
A likely scenario is that cuprate stripes at $x=\frac 1 8$ are bond-centered
(Fig.~\ref{fig:geom}) instead of site-centered,
as the latter situation would imply hole-poor regions with a width of three sites,
which then favor spin order closer to the semi-classical limit \cite{scheidl,erica}.
[Of course, more complicated ordered states, e.g., with strong dimerization realized
in a {\em site}-centered stripe structure, cannot be excluded.]

At doping levels different from $x=\frac 1 8$
a number of modifications can be considered:
Even in the absence of a modulation in the site-charge density
a bond-ordered (i.e. dimerized) state is still possible \cite{ssrmp,vs},
with a magnetic excitation spectrum very much similar to the higher-energy
part of our results.
For $x<\frac 1 8$, the stripe distance
will become larger upon decreasing doping \cite{waki}, which on the one hand
moves the ordering wavevector closer to $(\pi,\pi)$ -- this presumably
lowers the energy of the strong peak seen at $(\pi,\pi)$ \cite{scheidl,bala}.
On the other hand the hole-poor regions become wider, which stabilizes magnetic
order, thus reducing dimerization and quantum effects.
Therefore, for small doping one can expect the semi-classical spin-wave
approach \cite{scheidl,erica} to be more appropriate for the whole energy
range.
Finally, the present picture can in principle also apply to incommensurate
modulation periods, as seen in recent STM experiments \cite{ali},
e.g., if a superstructure like an array of period-4 and period-5 stripes
is realized.



We thank S.~Kivelson, S.~Scheidl, J.~Tranquada, G.~Uhrig, and in particular S.~Sachdev
for illuminating discussions.
This research was supported by the DFG
Center for Functional Nano\-structures Karls\-ruhe.



\begin{thebibliography}{}
\bibitem{param}
The precise location of the phase transition, i.e., $J_s^{cr}$, is not known
for the model in Fig.~\protect\ref{fig:geom}.
From the results for antiferromagnetic $J_s$ we can estimate
$J_s^{cr} \approx -0.2 ... -0.3 J$ for $J=J'$, which is not reproduced
by Eq. (\protect\ref{lameq}).
However, for the present calculation the most important aspect is to use
a value of $J_s$ which places the system (within our approximation)
close to the critical point -- this ensures the qualitatively correct
description of weak antiferromagnetism plus bond order.
\bibitem{error}
The accuracy of our approximation can roughly be estimated from Ref.~\protect\cite{eder}:
the errors in, e.g., the excitation energies are around 20\% of $J$.
\end{thebibliography}
\end{document}